\documentclass[aps,prc,nofootinbib,twocolumn,superscriptaddress,showpacs,
floatfix]{revtex4}
\usepackage{mathrsfs}
\usepackage{latexsym}
\usepackage{amsmath}
\usepackage{amssymb}
\usepackage{graphicx}
\usepackage{longtable}

\usepackage{bbm}
\usepackage{epsfig}
\usepackage[usenames]{color}
\usepackage[colorlinks=true,citecolor=blue,linkcolor=blue]{hyperref}

\begin{document}

%\begin{CJK*}{GBK}{song}

%\begin{CJK*}{GB}{}

\title{Phase diagrams in the Hadron-PNJL model}

\author{G.Y. Shao}
%\email[Corresponding author: ]{shaogy@pku.edu.cn}
\affiliation{INFN-Laboratori Nazionali del Sud, Via S. Sofia 62, I-95123
Catania, Italy}

\author{M. Di Toro}
\email[Corresponding author: ]{ditoro@lns.infn.it}
\affiliation{INFN-Laboratori Nazionali del Sud, Via S. Sofia 62, I-95123
Catania, Italy}
\affiliation{Physics and Astronomy Dept., University of Catania, Italy}

\author{V. Greco}
\affiliation{INFN-Laboratori Nazionali del Sud, Via S. Sofia 62, I-95123
Catania, Italy}
\affiliation{Physics and Astronomy Dept., University of Catania, Italy}

\author{M. Colonna}
\affiliation{INFN-Laboratori Nazionali del Sud, Via S. Sofia 62, I-95123
Catania, Italy}

\author{S. Plumari}
\affiliation{INFN-Laboratori Nazionali del Sud, Via S. Sofia 62, I-95123
Catania, Italy}
\affiliation{Physics and Astronomy Dept., University of Catania, Italy}

\author{B. Liu}
\affiliation{IHEP, Chinese Academy of Sciences, Beijing, 100049 China}
\affiliation{Theoretical Physics Center for Scientific Facilities, \\Chinese
Academy of Sciences, Beijing, 100049 China}

\author{Y.X. Liu}
\affiliation{Department of Physics and State Key Laboratory of \\
Nuclear Physics and Technology,
Peking University, Beijing 100871, China}
\affiliation{Center of Theoretical Nuclear Physics,\\ National Laboratory of
Heavy Ion Accelerator, Lanzhou 730000, China}

%\date{\today}

\begin{abstract}
The two-Equation of State (Two-EoS) model is used to describe the hadron-quark
phase transition in dense-hot matter formed  in heavy-ion collisions.
The non-linear Walecka model is used to describe the hadronic phase.
For the quark phase, the Nambu--Jona-Lasinio model coupled to Polyakov-Loop
fields (PNJL) is used to
include both the chiral and (de)confinement dynamics. The phase diagrams are
derived from the Gibbs conditions and
compared with the results obtained in the Hadron-NJL model without confinement.
As in the Hadron-NJL case a first order transition is observed, but with a
Critical-End-Point at much higher temperature, consequence of  the
confinement mechanism that reduces the degrees of freedom of the quark matter
in proximity of the phase transition.
Particular attention is devoted to  the phase transition in isospin
 asymmetric matter. Interesting isospin effects are found at high baryon
density and reduced temperatures, in fact common also to other quark models,
like MIT-Bag and NJL model. Some possible observation signals are suggested
to probe in Heavy-Ion Collision (HIC) experiments at intermediate energies.

\end{abstract}


\maketitle

\section{Introduction}

The exploration of the phase diagram of strongly interacting matter and
search for the signals of the phase transition from hadronic to quark-gluon
phase are very important in both
theory and experiment. As a fundamental tool, lattice QCD provides us the best
framework for investigation of non-perturbative phenomena such as confinement
and quark-gluon plasma formation at finite temperature and vanishing~(small)
chemical potential \cite{Karsch01, Karsch02, Allton02, Kaczmarek05,
Cheng06, YAoki99, Borsanyi10}. However,
lattice QCD suffers the serious problem of the fermion determinant with
three color at finite $\mu$.
Although several approximation methods have been taken to try to evade this
problem \cite{Fodor02,Fodor03,Elia09,Ejiri08,Clark07}, the validity of
lattice simulations at finite   chemical potential
is still limited to the region $\mu_q/T<1$~\cite{Fukushima11}. The results
 obtained with
$\mu_q/T>1$ should be taken with care.

On the other hand, many phenomenological models
 \cite{Nambu61,Toublan03,Werth05,Abuki06}, as well as the more
microscopic Dyson-Schwinger equations (DSEs) approach \cite{Qin11},
have been proposed to derive a complete description of  QCD phase diagram.
Among these effective models, the Nambu--Jona-Lasinio model~(NJL) is a
predominant one, since it offers a simple illustration of chiral symmetry
breaking and restoration, a
key feature of QCD~\cite{Volkov84,Hatsuda84,Klevansky92,Hatsuda94,
Alkofer96,Buballa05, Rehberg95}. Moreover, it provides a complicated
phase diagram of color superconductivity at high density~\cite{
Shovkovy03,Huang03,Alford08}. One deficiency of the standard NJL model is
that quarks
are not confined. Recently, an improved version of the NJL model coupled to
Polyakov-Loop fields  (PNJL) has been proposed~\cite{Fukushima04}.
The PNJL model takes into account both the chiral symmetry and
(de)confinement effect, giving a good interpretation of  lattice data at zero
chemical potential and finite temperature. At the same time it is able to make
 predictions in regions that cannot be presently reached in  lattice
calculation~\cite{Ratti06,Costa10,Schaefer10,Herbst11,Kashiwa08,Abuki08,Fu08}.

Most effective models, including the PNJL model, describe the
hadron--quark-gluon  phase transition
based on quark degrees of freedom. As a matter of fact, at low
temperature and small chemical potential,  QCD dynamics should be
governed by hadrons. Therefore, it is natural to describe the
strongly interacting matter with hadronic degrees of freedom at low $T$ and
small $\mu$ and quarks  at high $T$ and large $\mu$.
This picture can be easily realized following a two equation of state
(Two-EoS) model, where hadronic and quark phases are connected by the
Gibbs (Maxwell) criteria.
Such approach is widely used in describing the phase transition in the
interior of compact star in beta-equilibrium~(e.g.,\,\cite{Glendenning92,
Glendenning98,Burgio02,Maruyama07,Yang08,Shao10,Xu10} ). Recently, it  has
also been adopted to explore
the phase diagram of hadron-quark transition at finite temperature and density
related to Heavy-Ion Collision (HIC)
~\cite{Muller97, Toro06, Torohq09, Pagliara10, Cavagnoli10}.
Moreover, in these studies more attention was paid to isospin asymmetric 
matter, and
some
observable effects have been suggested to be seen in charged meson yield
ratio and on  the
onset of quark number scaling of the meson/baryon elliptic flows
in Ref.~\cite{Toro06, Torohq09}. Such heavy ion connection provides us a new
orientation to
investigate the hadron-quark phase transition, and it can stimulate some new
relevant researches in this field.

We have previously studied the hadron-quark phase transition in the
Two-EoS model by using the MIT-Bag model~\cite{Torohq09}
and  NJL model~\cite{Shao11} to describe  quark matter, respectively.
In particular a kind of Critical-End-Point (CEP) of a first order transition
has been found at about $T=80$ MeV and $\mu=900$ MeV when NJL model
is considered for the quark matter.
%to describe the quark phase. The calculations showed the hadron-quark
%phase transition highly
%depends the bag constant when the bag model is used for the quark phase,
%and only the phase diagram
%at low temperature and high density are obtained with NJL model.
In this paper, in order to obtain more reliable results and  predict possible
observables in the experiments, an improved calculation, within
the Two-EoS approach, has been performed.
We take the PNJL Lagrangian to describe the properties of quark matter, with
the interaction between quarks and Polyakov-Loop,
where both the chiral and (de)confinement dynamics are included simultaneously.
We are not considering here color pairing correlations,
that are affecting the isospin asymmetry  \cite{Pagliara10},
since in heavy ion
collisions the high density system will be always formed at rather large
temperatures \cite{Toro06}.

We obtain the phase diagrams of hadron--quark-gluon phase transition in $
T-\rho_{B}^{}$ and  $T-\mu_B^{}$ planes.
We compare the obtained results with those given in \cite{Shao11} where
the NJL model
is used to describe the quark phase.  The calculation shows that
the phase-transition curves are greatly
modified when both the chiral dynamics and (de)confinement effect are
considered, in particular in the high temperature and low chemical potential
region.
We still see a first order transition but the CEP is now at much higher
temperature and lower chemical potential. In fact the CEP temperature is much
closer
 to the critical temperature (for a crossover) given by lattice calculation at
vanishing chemical potential. Our results seem to stress the importance
of an extension of lattice calculations up to quark chemical potentials
around $\mu_q/T_c \simeq 1$.

In addition  we address the discussion about the non-coincidence of
chiral and deconfinement phase transition at large chemical potential 
and low temperature,  relevant to  the  formation of
 quarkyonic matter.
% and the possible influence to the phase transition in the Two-EoS model
%if the entanglement is considered between chiral condensate and
%Polyakov loop.%

Finally the calculation confirms that the onset density of hadron-quark phase
transition is much smaller  in  isospin asymmetric
than that of symmetric
matter, and therefore it will be more easy to  probe the mixed phase
in experiments.

The paper is organized as follows. In Section II, we describe
briefly the Two-EoS approach and give the relevant formulae of the
hadronic non-linear Walecka model
and the PNJL effective theory. In Section III, we discuss the
expected effects of the confinement dynamics. The quark matter phase transition
are presented in Section IV for the NJL as well as the PNJL models.
Section V is devoted to the 
phase diagrams within the Two-EoS frame to the comparison with the results
using only the pure quark PNJL model to describe both phases.
Moreover, we present some discussions and conclusions about the phase
transition, as well as some suggestions
for further study. Finally, a summary is given in Section VI.

\vskip 0.5cm

\section{ hadron matter, quark matter and the mixed phase}
%\allowdisplaybreaks

%\newpage
In our Two-EoS approach, the hadron matter and quark matter are
described by the non-linear Walecka model and by the PNJL model, respectively.
For the mixed phase between pure hadronic and quark matter,
the two phases are connected each other with the Gibbs conditions deduced
from thermal, chemical and mechanical equilibriums.
In this section, we will first give a short introduction of the nonlinear
Walecka model for the hadron matter
and the PNJL model for quark matter, then we construct the mixed phase
with the Gibbs criteria
based on baryon and isospin charge conservations during the transition.

For hadron phase, the non-linear Relativistic Mean Field (RMF) approach is
used, which provides an excellent description of  nuclear matter and finite
nuclei as well as of compressed matter properties probed with high energy HIC
 ~\cite{Muller97, Toro06, Torohq09, Liubo11, Toro09}.
The exchanged mesons include the isoscalar-scalar
meson $\sigma$ and  isoscalar-vector meson $\omega$ ($NL$ force, 
for isospin symmetric matter), 
isovector-vector
meson $\rho$  and  isovector-scalar meson $\delta$, 
($NL\rho$ and $NL\rho\delta$ forces, for isospin asymmetric matter). 

 The effective
Lagrangian is written as
\begin{widetext}
\begin{eqnarray}
\cal{L} &=&\bar{\psi}[i\gamma_{\mu}\partial^{\mu}- M
          +g_{\sigma }\sigma+g_{\delta }\boldsymbol\tau_{}
\cdot\boldsymbol\delta
          -g_{\omega }\gamma_{\mu}\omega^{\mu}
          -g_{\rho }\gamma_{\mu}\boldsymbol\tau_{}\cdot\boldsymbol
\rho^{\mu}]\psi
           \nonumber\\
   & &{}+\frac{1}{2}\left(\partial_{\mu}\sigma\partial^
{\mu}\sigma-m_{\sigma}^{2}\sigma^{2}\right)
       - \frac{1}{3} b\,(g_{\sigma} \sigma)^3-\frac{1}{4} c\,
(g_{\sigma} \sigma)^4
          +\frac{1}{2}\left(\partial_{\mu}\delta\partial^{\mu}\delta
          -m_{\delta}^{2}\delta^{2}\right)  \nonumber\\
       & &{}+\frac{1}{2}m^{2}_{\omega} \omega_{\mu}\omega^{\mu}
          -\frac{1}{4}\omega_{\mu\nu}\omega^{\mu\nu}
          +\frac{1}{2}m^{2}_{\rho}\boldsymbol\rho_{\mu}\cdot\boldsymbol
\rho^{\mu}
          -\frac{1}{4}\boldsymbol\rho_{\mu\nu}\cdot\boldsymbol\rho^{\mu\nu},
 \end{eqnarray}
\end{widetext}
where the antisymmetric tensors of vector mesons are given by
\begin{equation}
\omega_{\mu\nu}= \partial_\mu \omega_\nu - \partial_\nu
\omega_\mu,\qquad \nonumber \rho_{\mu\nu} \equiv\partial_\mu
\boldsymbol\rho_\nu -\partial_\nu \boldsymbol\rho_\mu.
\end{equation}

The nucleon chemical potential and effective mass in nuclear medium can be expressed as
\begin{equation}
\mu_{i}^{} =\mu_{i}^{*}+g_{\omega
}\omega+g_{\rho}\tau_{3i}^{}\rho \, ,
\end{equation}
and
\begin{equation}
M_{i}^{*} = M -g_{\sigma }^{}\sigma-g_{\delta
}^{} \tau_{3i}^{} \delta,
\end{equation}
where $M$ is the free nucleon mass, $ \tau_{3p}^{}=1$ for proton and
$\tau_{3n}^{}=-1$ for neutron, and $\mu_{i}^{*}$
is the effective chemical potential which reduces to Fermi energy
$E_{Fi}^{*}=\sqrt{k_{F}^{i^{2}}+M_{i}^{*^{2}}}$ at zero temperature.
The baryon and isospin chemical potentials in the hadron phase are defined as
\begin{equation}
\mu_{B}^{H} =\frac{\mu_{p}+\mu_{n}}{2},\ \ \ \ \ \mu_{3}^{H} =\frac
{\mu_{p}-\mu_{n}}{2}.
\end{equation}
The energy density and pressure of nuclear matter at finite temperature are
derived as
\begin{widetext}
\begin{equation}
\varepsilon^{H} =  \sum_{i=p,n}\frac{2}{(2\pi)^3} \int \! d^3
\boldsymbol k \sqrt{k^2 + {M^*_i}^2}(f_{i}(k)+\bar{f}_{i}(k))+
\frac{1}{2}m_\sigma^2 \sigma^2 +
\frac{b}{3}\,(g_{\sigma }^{} \sigma)^3+ \frac{c}{4}\,(g_{\sigma
}^{} \sigma)^4
+\frac{1}{2}m_\delta^2 \delta^2+\frac{1}{2}m_\omega^2 \omega^2 +
\frac{1}{2}m_\rho^2 \rho^2 \, ,
\end{equation}
\begin{equation}
P^{H}  =  \sum_{i=p,n} \frac{1}{3}\frac{2}{(2\pi)^3} \int \!
d^3 \boldsymbol k \frac{k^2}{\sqrt{k^2 + {M^*_i}^2}}(f_{i}(k)+
\bar{f}_{i}(k))- \frac{1}{2}m_\sigma^2
\sigma^2 - \frac{b}{3}\,(g_{\sigma }^{} \sigma)^3 -
\frac{c}{4}\,(g_{\sigma }^{} \sigma)^4
-\frac{1}{2}m_\delta^2 \delta^2+\frac{1}{2}m_\omega^2 \omega^2 +
\frac{1}{2}m_\rho^2 \rho^2 \, .
\end{equation}
\end{widetext}
where $f_{i}(k)$ and $\bar{f}_{i}(k)$ are the fermion and antifermion
distribution functions
for proton and neutron ($i=p,\,n$):
\begin{equation}
  f_{i}(k)=\frac{1}{1+\texttt{exp}\{(E_{i}^{*}(k)-\mu_{i}^{*})/T\}} ,
\end{equation}
\begin{equation}
  \bar{f}_{i}(k)=\frac{1}{1+\texttt{exp}\{(E_{i}^{*}(k)+\mu_{i}^{*})/T\}}.
\end{equation}
The effective chemical potentials $\mu_{i}^{*}$ are determined by the nucleon
densities
\begin{equation}\label{hadrondensity}
  \rho_i= 2 \int \! \frac{d^3 \boldsymbol k}{(2\pi)^{3}}( f_{i}(k)-
\bar{f}_{i}(k)).
\end{equation}
With the baryon number density $\rho=\rho_{B}^{H}=\rho_p+\rho_n$ and isospin
density $\rho_{3}^{H}=\rho_p-\rho_n$.
The asymmetry parameter can be defined as
\begin{equation}
   \alpha^{H}\equiv-\frac{\rho_{3}^{H}}{\rho_{B}^{H}}=\frac{\rho_n-
\rho_p}{\rho_p+\rho_n}.
\end{equation}

In this study the parameter set $NL\rho\delta$~\cite{Toro06}
will be used to describe the properties of hadron matter.
The model parameters is determined by
calibrating the properties of symmetric nuclear matter at zero temperature
and normal
nuclear density. Our parameterizations are also tuned to reproduce collective
flows and
particle production at higher energies, where some hot and dense matter
is probed, see~\cite{Toro09} and refs. therein.

We take the PNJL model to describe the quark matter.
In the pure gauge theory,  the Polyakov-Loop serves as an order parameter
for the $\mathbb{Z}_3$ symmetry breaking transition from low to high
temperature, i.e. for the  transition from a confined to a deconfined phase.
In the real world quarks are coupled to the Polyakov-Loop, which explicitly
breaks the
$\mathbb{Z}_3$ symmetry. No rigorous order parameter is established for
the deconfinement
phase transition. However, the Polyakov loop can still be practicable
to distinguish
a confined phase from a deconfined one.

The Lagrangian density in the three-flavor PNJL model is taken as
\begin{eqnarray}\label{polylagr}
\mathcal{L}_{q}&=&\bar{q}(i\gamma^{\mu}D_{\mu}-\hat{m}_{0})q+
G\sum_{k=0}^{8}\bigg[(\bar{q}\lambda_{k}q)^{2}+
(\bar{q}i\gamma_{5}\lambda_{k}q)^{2}\bigg]\nonumber \\
           &&-K\bigg[\texttt{det}_{f}(\bar{q}(1+\gamma_{5})q)+\texttt{det}_{f}
(\bar{q}(1-\gamma_{5})q)\bigg]\nonumber \\ \nonumber \\
&&-\mathcal{U}(\Phi[A],\bar{\Phi}[A],T),
\end{eqnarray}
where $q$ denotes the quark fields with three flavors, $u,\ d$, and
$s$, and three colors; $\hat{m}_{0}=\texttt{diag}(m_{u},\ m_{d},\
m_{s})$ in flavor space; $G$ and $K$ are the four-point and
six-point interacting constants, respectively. The four-point
interaction term in the Lagrangian keeps the $SU_{V}(3)\times
SU_{A}(3)\times U_{V}(1)\times U_{A}(1)$ symmetry, while the 't
Hooft six-point interaction term breaks the $U_{A}(1)$ symmetry.

The covariant derivative in the Lagrangian density is defined as
 $D_\mu=\partial_\mu-iA_\mu$.
The gluon background field $A_\mu=\delta_\mu^0A_0$ is supposed to 
be homogeneous
and static, with  $A_0=g\mathcal{A}_0^\alpha \frac{\lambda^\alpha}{2}$, where
$\frac{\lambda^\alpha}{2}$ are $SU(3)$ color generators.
The effective potential $\mathcal{U}(\Phi[A],\bar{\Phi}[A],T)$ is expressed in terms of the traced Polyakov loop
$\Phi=(\mathrm{Tr}_c L)/N_C$ and its conjugate
$\bar{\Phi}=(\mathrm{Tr}_c L^\dag)/N_C$. The Polyakov loop $L$  is a matrix in color space
\begin{equation}
   L(\vec{x})=\mathcal{P} exp\bigg[i\int_0^\beta d\tau A_4 (\vec{x},\tau)   \bigg],
\end{equation}
where $\beta=1/T$ is the inverse of temperature and $A_4=iA_0$.

The Polyakov loop can be expressed in a more intuitive physical form
as
\begin{equation}\label{poly}
\Phi = \exp{[-\beta F_q(\vec{x})]}
\end{equation}
where $F_q$ is the free energy required to add an isolated quark to the system.
So it will go from zero in the confined phase up to a finite value when
deconfinement is reached \cite{McL81}.

Different effective potentials are
adopted in the literature~\cite{Ratti06,Robner07,Fukushima08}.
The logarithmic one given in~\cite{Robner07} will be used in our calculation, which can
reproduce well the data obtained in lattice calculation.
The corresponding effective potential reads
%\begin{widetext}
%\begin{equation}
%     \frac{\mathcal{U}(\Phi,\bar{\Phi},T)}{T^4}=-\frac{a(T)}{2}\bar{\Phi}\Phi+b(T)\mathrm{ln}[1-6\bar{\Phi}\Phi+
%         4(\bar{\Phi}^3+\Phi^3)-3(\bar{\Phi}\Phi)^2]
%\end{equation}
%\end{widetext}
%
\begin{eqnarray}
     \frac{\mathcal{U}(\Phi,\bar{\Phi},T)}{T^4}&=&-\frac{a(T)}{2}\bar{\Phi}\Phi \\
                                                &&+b(T)\mathrm{ln}\bigg[1-6\bar{\Phi}\Phi+4(\bar{\Phi}^3+\Phi^3)-3(\bar{\Phi}\Phi)^2\bigg], \nonumber
\end{eqnarray}
where
\begin{equation}
    a(T)=a_0+a_1\bigg(\frac{T_0}{T}\bigg)+a_2\bigg(\frac{T_0}{T}\bigg)^2,
\end{equation}
and
\begin{equation}
    b(T)=b_3\bigg(\frac{T_0}{T}\bigg)^3.
\end{equation}

We note that in this version of PNJL the direct coupling between quark 
condensates 
and  Polyakov loop is only via the covariant derivative in the Lagrangian
density Eq.(\ref{polylagr}).

The parameters $a_i$, $b_i$ are precisely fitted according to the lattice
result of  QCD thermodynamics in
pure gauge sector. $T_0$ is found to be 270 MeV as the critical
temperature for the deconfinement phase transition of the gluon part at zero
chemical potential~\cite{Fukugita90}. When fermion fields are included,
a rescaling of $T_0$
is usually implemented to obtain consistent result between model calculation
and full lattice simulation which gives a critical phase-transition temperature
$T_c=173\pm8$ MeV~\cite{Karsch01,Karsch02,Kaczmarek05}. In this study we
rescale $T_0=210$\, Mev so as to produce $T_c=171$ MeV for the
phase transition temperature at zero chemical potential.

In the mean field approximation, quarks can be seen as free quasiparticles
with constituent masses $M_i$,
and the dynamical quark masses~(gap equations) are obtained as
\begin{equation}
M_{i}=m_{i}-4G\phi_i+2K\phi_j\phi_k\ \ \ \ \ \ (i\neq j\neq k),
\label{mass}
\end{equation}
with $i~=~u, d, s$,
where $\phi_i$ stands for the quark condensate.
%The corresponding propagator in the background field $A_4$ is
%\begin{equation}
%     S_i(p)=-(p\!\!\slash-M_i+\gamma_0(\mu-iA_4))^{-1}.
%\end{equation}
The thermodynamic potential of the PNJL model at the mean field level is
expressed as
\begin{eqnarray}
\Omega&=&\mathcal{U}(\bar{\Phi}, \Phi, T)+2G\left({\phi_{u}}^{2}
+{\phi_{d}}^{2}+{\phi_{s}}^{2}\right)-4K\phi_{u}\,\phi_{d}\,\phi_{s}\nonumber \\
&&-T \sum_n\int \frac{\mathrm{d}^{3}p}{(2\pi)^{3}}\mathrm{Trln}\frac{S_i^{-1}(i\omega_n,\vec{p})}{T}.
\end{eqnarray}
Here $S_i^{-1}(p)=-(p\!\!\slash-M_i+\gamma_0(\mu_i-iA_4))$, with 
$\mu_i$ quark chemical potential, is the inverse fermion propagator in 
the background field $A_4$,
and the trace has to be taken in color, flavor, and Dirac space.
After summing over the fermion  Matsubara frequencies, $p^0=i\omega_n=(2n+1)\pi T$,
the thermodynamic potential can be written as
\begin{widetext}
\begin{eqnarray}
\Omega&=&\mathcal{U}(\bar{\Phi}, \Phi, T)+2G\left({\phi_{u}}^{2}
+{\phi_{d}}^{2}+{\phi_{s}}^{2}\right)-4K\phi_{u}\,\phi_{d}\,\phi_{s}-2\int_\Lambda \frac{\mathrm{d}^{3}p}{(2\pi)^{3}}3(E_u+E_d+E_s) \nonumber \\
&&-2T \sum_{u,d,s}\int \frac{\mathrm{d}^{3}p}{(2\pi)^{3}} \bigg[\mathrm{ln}(1+3\Phi e^{-(E_i-\mu_i)/T}+3\bar{\Phi} e^{-2(E_i-\mu_i)/T}+e^{-3(E_i-\mu_i)/T}) \bigg]\nonumber \\
&&-2T \sum_{u,d,s}\int \frac{\mathrm{d}^{3}p}{(2\pi)^{3}} \bigg[\mathrm{ln}(1+3\bar{\Phi} e^{-(E_i+\mu_i)/T}+3\Phi e^{-2(E_i+\mu_i)/T}+e^{-3(E_i+\mu_i)/T}) \bigg],
\end{eqnarray}
\end{widetext}
where $E_i=\sqrt{\vec{p}^{\,2}+M_i^2}$ is the energy of quark flavor $i$.

We remark some interesting differences with respect to the thermodynamical
potential derived within the pure NJL model, see \cite{Buballa05,Shao11}.
Apart the presence of  the effective potential  
$\mathcal{U}(\bar{\Phi}, \Phi, T)$,  the Polyakov
 loop is mostly acting on the quark-antiquark distribution functions, in the 
direction of a reduction, on the way to confinement. This is largely modifying
the quark pressure, as seen in the calculations. Moreover, in spite of
the minimal coupling introduced in the Lagrangian Eq.(\ref{polylagr}),
 only in the covariant derivative, 
the quark condensates will be strongly affected by the Polyakov loop via
the modified $q,~\bar{q}$ distribution functions. This will be also clearly 
observed in the comparison of NJL and PNJL phase diagrams.

The values of $\phi_u, \phi_d, \phi_s, \Phi$ and $\bar{\Phi}$ are determined by minimizing the thermodynamical
potential
\begin{equation}
\frac{\partial\Omega}{\phi_u}=\frac{\partial\Omega}{\phi_d}=\frac{\partial\Omega}{\phi_s}=\frac{\partial\Omega}{\Phi}=\frac{\partial\Omega}{\bar\Phi}=0.
\end{equation}
All the thermodynamic quantities relevant to the bulk properties of quark matter can be obtained from $\Omega$. Especially, the pressure
and energy density should be zero in the vacuum.

The baron (isospin) density and baryon (isospin )chemical potential in quark
phase are defined as follows
\begin{equation}
\rho_{B}^Q=\frac{1}{3}(\rho_u+\rho_d),\ \ \ \ \rho_{3}^Q=\rho_u-\rho_d,
\end{equation}
and
\begin{equation}
\mu_{B}^Q=\frac{3}{2}(\mu_u+\mu_d),\ \ \ \ \mu_{3}^Q=\frac{1}{2}(\mu_u-\mu_d).
\end{equation}
The corresponding asymmetry parameter of quark phase is defined as
\begin{equation}
   \alpha^{Q}\equiv-\frac{\rho_{3}^{Q}}{\rho_{B}^{Q}}=3\frac{\rho_d-\rho_u}
{\rho_u+\rho_d}.
\end{equation}

As an effective model, the (P)NJL model is not
renormalizeable, so a cut-off $\Lambda$ is implemented in 3-momentum
space for divergent integrations. We take the model parameters:
$\Lambda=603.2$ MeV, $G\Lambda^{2}=1.835$, $K\Lambda^{5}=12.36$,
$m_{u,d}=5.5$  and $m_{s}=140.7$ MeV, determined
by fitting $f_{\pi},\ M_{\pi},\ m_{K}$ and $\ m_{\eta}$ to their
experimental values~\cite{Rehberg95}. The coefficients in Polyakov effective potential
are listed in Table \ref{tab:1}.
\begin{table}[ht]
\tabcolsep 0pt \caption{\label{tab:1}Parameters in Polyakov effective potential given in~\cite{Robner07}}
\setlength{\tabcolsep}{2.5pt}
\begin{center}
\def\temptablewidth{0.58\textwidth}
%{\rule{\temptablewidth}{0.5pt}}
\begin{tabular}{c c c c}
\hline
\hline
   {$a_0$}                      & $a_1$        & $a_2$      & $a_3$           \\  \hline
   $ 3.51$                   & -2.47        &  15.2      & -1.75               \\ \hline
\hline
\end{tabular}
 % {\rule{\temptablewidth}{0.5pt}}
\end{center}
\end{table}

So far we have introduced how to describe the hadronic  and quark phase by
the RMF hadron model and PNJL quark model, respectively.
The key point in the Two-EoS model is to construct the phase transition
from  hadronic  to quark matter.
As mentioned above, the two phases are connected by Gibbs criteria, i.e.,
the thermal, chemical and mechanical equilibrations
being required. For the Hadron--quark-gluon phase transition relevant to
heavy-ion collision of duration  about
$10^{-22}sec$, ($10~-~20~fm/c$), thermal equilibration is only possible
for  strongly interacting processes, where baryon number and isospin
conservations are preserved.
So the strange-antistrange quark number may be rich, but the net strange
quark number should be zero  before the beginning
of hadronization in the  expansion stage~\cite{Greiner87}, which can be
approximately realized by requiring $\mu_s=0$~(Hadronization is out of the
range of this study ).

Based on the conservations of  baryon number and
isospin during strong interaction, the  Gibbs conditions describing
the phase transition
can be expressed by
\begin{eqnarray}
& &\mu_B^H(\rho_B^{},\rho_3^{},T)=\mu_B^Q(\rho_B^{},\rho_3^{},T)\nonumber\\
& &\mu_3^H(\rho_B^{},\rho_3^{},T)=\mu_3^Q(\rho_B^{},\rho_3^{},T)\nonumber\\
& &P^H(\rho_B^{},\rho_3^{},T)=P^Q(\rho_B^{},\rho_3^{},T),
\end{eqnarray}
where   $\rho_B^{}=(1-\chi)\rho_B^{H}+\chi \rho_B^{Q}$ and
$\rho_3^{}=(1-\chi)\rho_3^{H}+\chi \rho_3^{Q}$ are the total
baryon density, isospin density  of the mixed phase, respectively,
 and $\chi$ is the fraction of quark matter.
The global asymmetry parameter $\alpha$ for
the mixed phase is
\begin{equation}
   \alpha\equiv-\frac{\rho_{3}^{}}{\rho_{B}^{}}= \frac{(1-\chi)\rho_3^{H}+
\chi \rho_3^{Q}}{(1-\chi)\rho_B^{H}+\chi \rho_B^{Q}}=\alpha^H\mid_{\chi=0}^{} =  \alpha^Q\mid_{\chi=1}^{},
\end{equation}
which is determined by the heavy-ion source formed in experiments.

\begin{figure}[htbp]
\begin{center}
\includegraphics[scale=0.29]{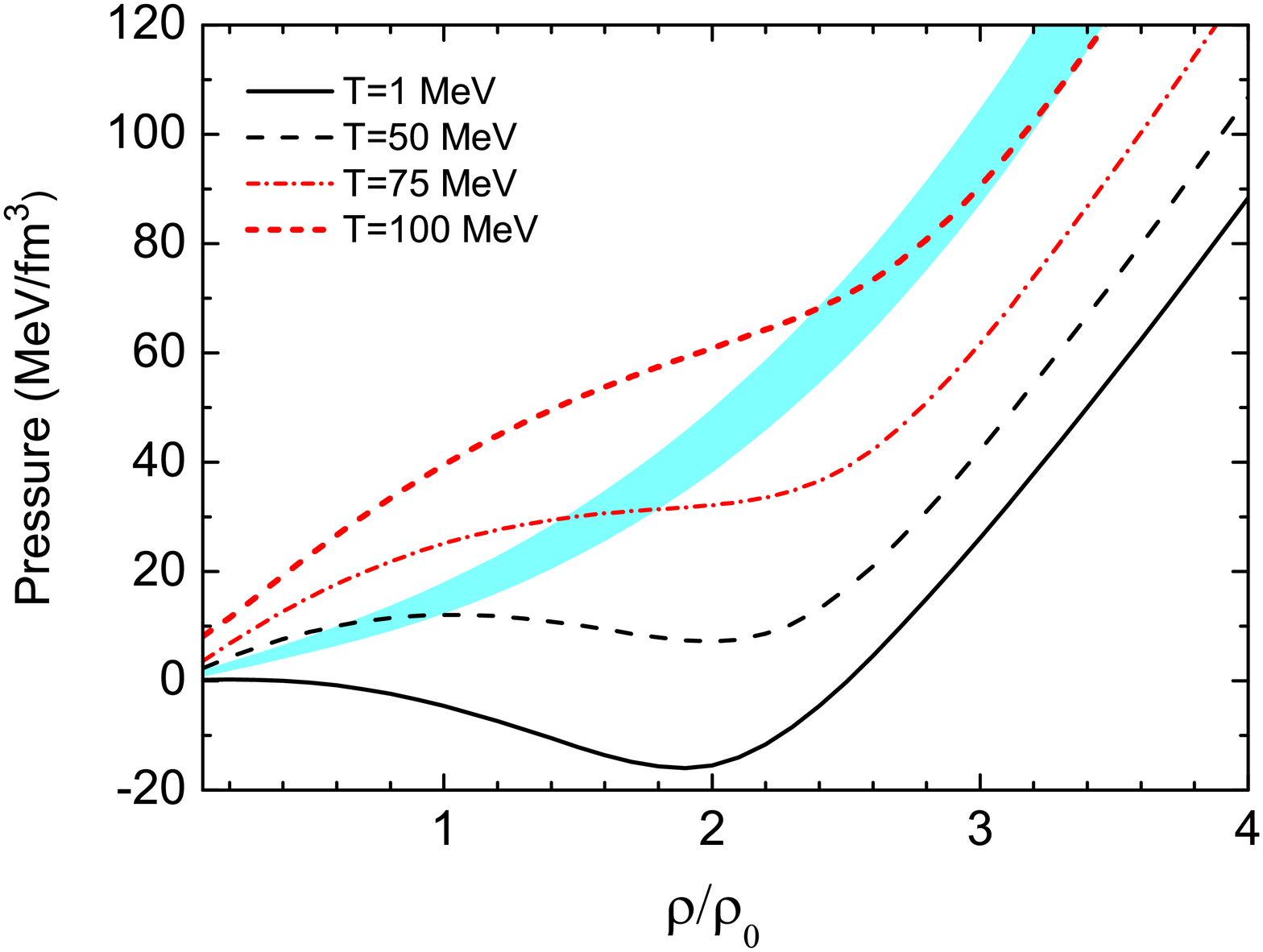}
\caption{\label{fig:Pressure-NJL}Pressure of quark matter as function of 
baryon
density at different temperatures in the NJL model.Isospin symmetric matter.
In the shaded area we show also the Hadron (NL) curves in the temperature 
region between 75 and 100 MeV.}
\end{center}
\end{figure}

\begin{figure}[htbp]
\begin{center}
\includegraphics[scale=0.29]{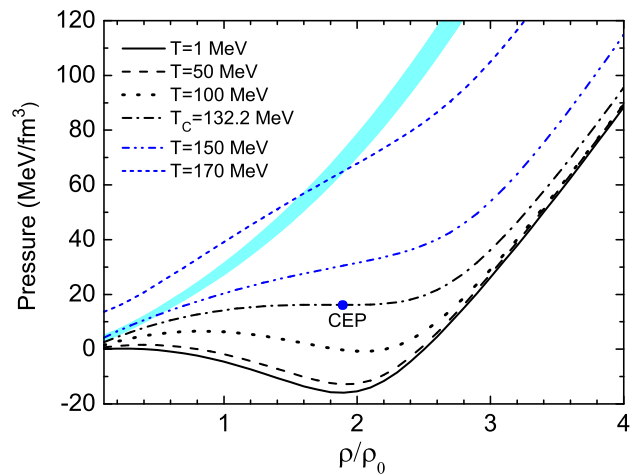}
\caption{\label{fig:Pressure-PNJL}Pressure of quark matter as function of 
baryon density at different temperatures in the PNJL model.
Isospin symmetric matter.
In the shaded area we show also the Hadron (NL) curves in the temperature 
region between 150 and 170 MeV.}
\end{center}
\end{figure}

\section{Expected effect of the confinement dynamics}

Before showing detailed phase diagram results within the Two-EoS approach
it is very instructive to analyze
 the effects of
chiral and (de)confinement dynamics in the pure quark sector.
In order to understand the physics which is behind we will show separately the
results in the NJL, with the same parameters given before, and the PNJL
model, for isospin symmetric matter.
In Figs.~\ref{fig:Pressure-NJL},~\ref{fig:Pressure-PNJL} we plot the
pressure of isospin symmetric quark matter as function of baryon density for
different temperatures respectively for the NJL and  PNJL models.
In this calculation isospin symmetric matter with $\mu=\mu_u=\mu_d$
and $\mu_s=0$ is considered and $\phi_l$ stands for the chiral condensate 
of $u, d$ quarks.

From the two figures, we can see that the pressure has a local maximum  and a 
local minimum at
low temperature. The local extrema will  disappear with the increase of 
temperature. The 
temperature  with the disappearance of the two local extrema corresponds
to the  Critical-End-Point $CEP$ of the first order chiral transition, for 
a more detailed discussion 
please refer to \cite{Buballa05,Costa10}. It is interesting to note that
the critical temperature of the chiral transition is rather different in the
two cases, around 70 MeV in the NJL and around 130 MeV in the PNJL, while the 
density region is not much affected. This is due to the fact
 that for a fixed baryon density (or chemical potential) the NJL presents a 
much larger pressure for a given temperature, as clearly seen from the two
figures \ref{fig:Pressure-NJL},~\ref{fig:Pressure-PNJL} \cite{Hansen07}.
 This is a nice indication that when we have a 
coupling to the 
deconfinement, even if in the minimal way included here, the quark pressure 
at finite temperatures is reduced
since the quarks degrees of freedom start to decrease.

All that will imply important differences at higher temperatures since above 
the chiral restoration the quark pressure will rapidly increase reaching an 
end-point in the Two-EoS approach
where the matching to the hadron pressure will not be possible.
This will happen in different points of the ($T,\mu$), ($T,\rho$) planes for
the Hadron-NJL \cite{Shao11} and the Hadron-PNJL, and higher temperatures 
will be requested 
in the PNJL case.
In fact this can be also clearly seen from
Figs.~\ref{fig:Pressure-NJL} and ~\ref{fig:Pressure-PNJL}, of the NJL-
and PNJL-pressures, where we plot also the corresponding curves of the
hadronic EoS in the end-point regions (shaded area). The Gibbs (Maxwell)
conditions
have no solution with decreasing density
(chemical potential) and increasing temperature if we encounter a crossing
of the hadron and quark curves in the $T-\rho_B$ ($T-\mu_B$) plane, with
the quark pressure becoming larger than the hadron one.
We see that this is happening  for $T \simeq 75$~MeV and
$\rho/\rho_0 \simeq 1.8$ in the NJL case and for $T \simeq 170$~MeV and
$\rho/\rho_0 \simeq 1.6$ in the PNJL quark picture.
In conclusion, due to the noticeable quark pressure difference at finite 
temperatures, besides the Critical-End-Points, we expect in general
rather different phase diagrams  
given by the Hadron-NJL  
and Hadron-PNJL models. This will be seen in the Section V,
  Figs.~\ref{fig:T-Mu-with-delta-alpha=0} and
 \ref{fig:T-Mu-with-delta-alpha=02}.

\begin{figure}[htbp]
\begin{center}
\includegraphics[scale=0.3]{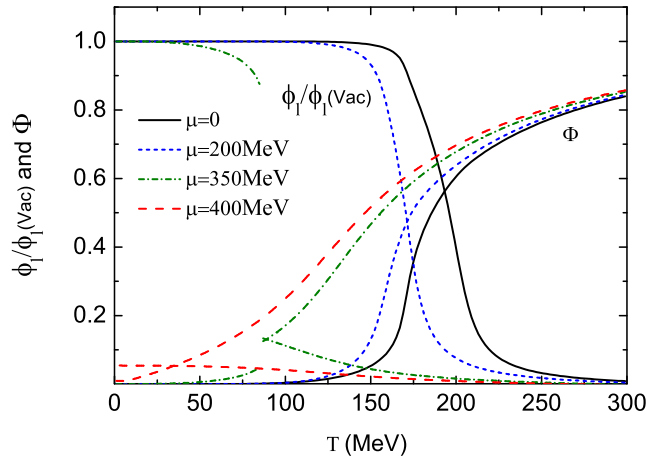}
\caption{\label{fig:phil-Phi-T}
Chiral condensate $\phi_l$ (normalized to the vacuum value) and Polyakov-Loop
$\Phi$ as functions
of temperature for various values of the quark chemical potential.
 Isospin symmetric matter.}
\end{center}
\end{figure}

\section{PNJL Phase diagram in the quark sector}

 In  order to better understand the effects of the coupling 
between quark 
condensates and Polyakov loop and also to compare with the Two-EoS results,
we discuss here also the  Phase diagram in the pure quark sector
obtained from the PNJL model.

We plot in Fig.~\ref{fig:phil-Phi-T} the temperature evolution of the chiral
condensate
$\phi_l$ and the  Polyakov-Loop $\Phi$ for various values of the quark chemical
potential.  $\Phi$ and $\bar{\Phi}$ have the same values
at zero chemical potential and their difference is very small at finite
chemical potential, hence
we only present the results of $\Phi$ in Fig.~\ref{fig:phil-Phi-T} and  later
in the discussion.

Firstly, we can see 
that the chiral condensate and Polyakov loop $\Phi$ vary continuously at
$\mu$~=~0 and 200 MeV, and there exist
sharp decreases (increases) at high temperature indicating the onset
of chiral (deconfinement) phase transitions.
These characteristics show that the corresponding chiral
and deconfinement phase transitions are crossovers for small chemical potential
at high temperature. At variance, for large chemical potentials, e.g.,
$\mu=350$~MeV,
the chiral condensate varies discontinuously  with the temperature, which
indicates the presence of a first order chiral phase transition, as already
seen in the pure NJL approach, although at much lower temperature
\cite{Buballa05}, as discussed in the previous Section.
The Polyakov loop is always showing a continuous behavior indicating that
we have only crossover transitions. The jump observed for the dash-dotted curve
corresponding to a $\mu=350$~MeV chemical potential is just an effect
of the coupling to the sharp variation of the quark condensates at the first 
order
chiral transition. Moreover this is happening in a region of very small values 
of the $\Phi$ field at lower temperatures. As a matter of fact such 
discontinuity disappears for the results at $\mu= 400$~MeV, i.e. above 
the chiral transition, see the dashed curves for both $\phi$ and $\Phi$ fields.

\begin{figure}[htbp]
\begin{center}
\includegraphics[scale=0.3]{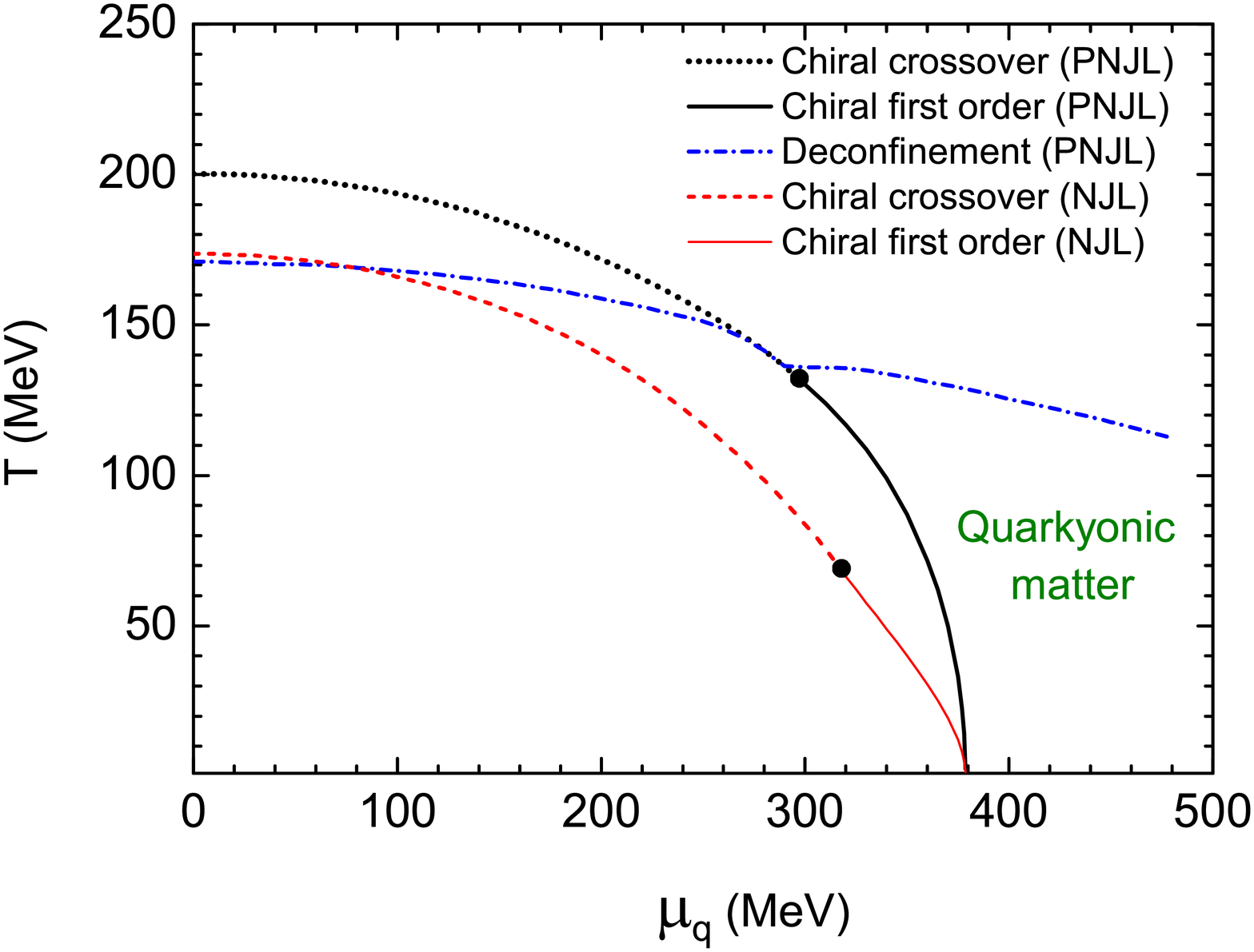}
\caption{\label{fig:Phase-Transition-PNJ} Phase diagram of the PNJL model.
The corresponding chiral phase transition for the pure NJL model is also
shown. Isospin symmetric matter.}
\end{center}
\end{figure}

Finally in Fig.~\ref{fig:Phase-Transition-PNJ}
we plot the phase diagram of the PNJL model in $T-\mu_q$ plane
(always for isospin symmetric matter). The phase transition curves are
obtained by requiring ${\partial \phi_l}/{\partial T}$ and
${\partial \Phi}/{\partial T}$ taking the maximum values.
For deconfinement phase transition the use of the maximum value of
${\partial \Phi}/{\partial T}$ as the phase-transition tracer is a
good choice when $\mu$ is not too large. In fact we see from
Fig.~\ref{fig:phil-Phi-T} a sharp increase of  $\Phi$
for $\mu=0$ and 200 MeV.
   However, with the increase of chemical potential,
although we are still able to observe the maximum of
${\partial \Phi}/{\partial T}$,
the width of the maximum increases. The peaks of
${\partial \Phi}/{\partial T}$ are more smoothed and this will be not any more
a well defined phase-transition parameter as $\mu$ is large.
Therefore, some authors take $\Phi=1/2$ as the phase transition parameter
\cite{Fukushima08,Sakai10,Sakai11}.
%Anyway similar phase-transition curves will be obtained.
In conclusion from Fig.~\ref{fig:phil-Phi-T}
we can see that the chiral phase transition
is  continuous at high temperature and relatively smaller chemical
potential, while  a first order phase transition takes place
at low temperature and larger chemical potential.
The {\it Critical-End-Point} ($CEP$) of the chiral transition appears at
$(132.2,~296.6)$~MeV in the $T-\mu_q$ plane, in agreement with similar
calculations \cite{Sakai10}.
% just as
%shown in Fig.~\ref{fig:phil-Phi-T}.
At variance, the deconfinement phase
transition is always a continuous crossover in the
PNJL model, but the peak of ${\partial \Phi}/{\partial T}$ becomes more
and more
smooth with the increase of baryon chemical potential. In addition, at large
chemical potential, a chirally restored but still
confined matter, the \emph{quarkyonic matter}, can be realized in the PNJL
model.
All that is reported in  Fig.~\ref{fig:Phase-Transition-PNJ} where we plot
the full phase diagram of the PNJL approach.

Here we give a short discussion about the coincidence of chiral and
deconfinement phase transitions as well as the presence of quarkyonic matter.
The temperature dependence of the chiral condensate and of the Polyakov
loop of Fig.~\ref{fig:phil-Phi-T} as well the PNJL phase diagram
of Fig. \ref{fig:Phase-Transition-PNJ} are obtained with the
rescaled parameter $T_0=210$~MeV.
The coincidence of chiral restoration and deconfinement takes place at
about $\mu_q=290$ MeV. If we take $T_0=270$~MeV, the approximate coincidence,
with the different phase-transition temperatures less than 10 MeV, will move
down to $\mu\simeq0$.  In any case,
there is only one cross point of the two phase transitions.
Up to now, the relation between
chiral-symmetry restoration and deconfinement phase transition is still an
open question. It is possible that the coincidence of chiral and deconfinement
phase
transition takes place in a wider range of chemical potentials. Such 
coincidence indeed has been
recently realized  by considering a larger coupling ($entanglement$)
between chiral condensate and Polyakov loop, with an explicit 
$\Phi$-dependence of the
condensate couplig $G(\Phi)$ and a chemical potential
dependent $T_0$ ~\cite{Sakai10,Sakai11}.

In the same Fig.~\ref{fig:Phase-Transition-PNJ} we report also the chiral 
transition curve for the pure NJL model (same parameters).
We note the the coupling between the chiral condensates and the Polyakov-Loop
fields ($\Phi,~\bar{\Phi}$) is mostly affecting the temperature of the 
chiral $CEP$ as expected from the pressure discussion of the previous Section.

From Fig.~\ref{fig:Phase-Transition-PNJ}
we can see that the deconfinement phase-transition temperature is
still high at large chemical potential, and so
the region of quarkyonic matter appears very wide. On the other hand the 
signature of a deconfinement transition is disappearing at large chemical 
potentials and lower temperatures.  Because of the lack of lattice
QCD data at large real chemical potentials,
more investigations are needed to study the physics in this range. The results
in \cite{Sakai11} also show that the range of quarkyonic matter shrinks
when  a $\mu$-dependent $T_0$ and/or a larger entanglement
between quark condensate and Polyakov loop is considered.

We remark that this ($T-\mu$) zone just represents the 
nuclear metter phase diagram region possibly reached in  the 
collision of 
heavy ions at intermediate energies and so it is of large interest to
perform Two-EoS predictions, wich should have a good connection to the
more fundamental results of effective quark models.
This is the subject of the next Section.

\section{Hadron-Quark Phase Transitions}

In the following  we will discuss the phase diagrams obtained in the
Two-EoS model, i.e., explicitly considering a hadronic EoS with the parameter
 set of $NL$
for symmetric matter and 
$NL\rho \delta$ for asymmetric matter at low density and chemical 
potential~\cite{BaranPR,Toro06}.

\begin{figure}[htbp]
\begin{center}
\includegraphics[scale=0.3]{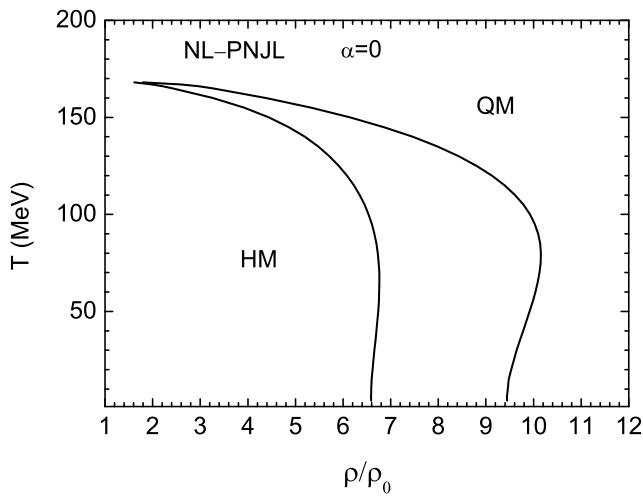}
\caption{\label{fig:T-RHO-with-delta-alpha=0}Phase diagram in  $T-\rho_B^{}$
plane in the Two-EoS model for symmetric matter. }
\end{center}
\end{figure}
 
\begin{figure}[htbp]
\begin{center}
\includegraphics[scale=0.3]{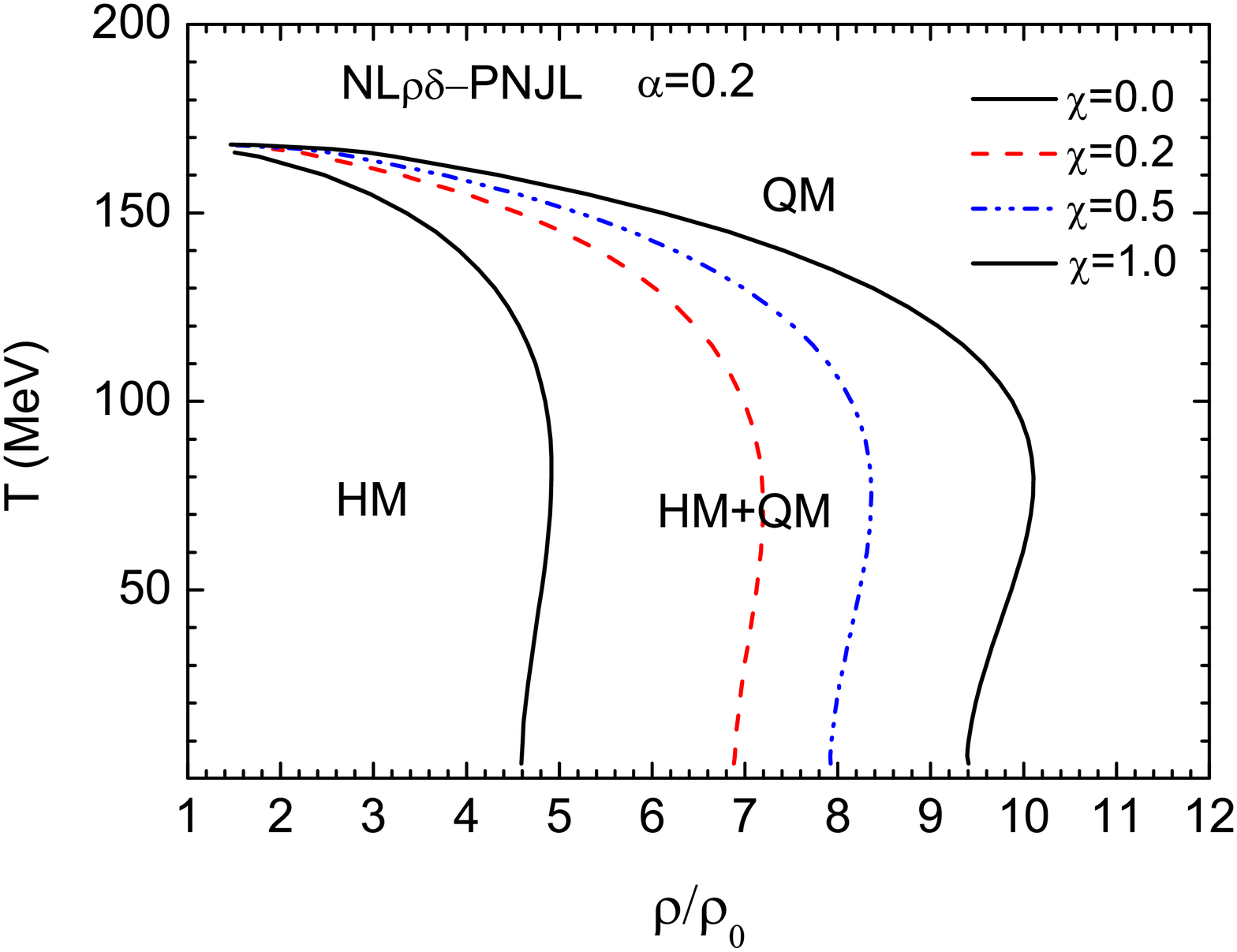}
\caption{\label{fig:T-RHO-with-delta-alpha=02}Phase diagram in  $T-\rho_B^{}$ plane in the
Two-EoS model for asymmetric matter with the global asymmetry parameter $\alpha=0.2$. $\chi$ represents
the fraction of quark matter.}
\end{center}
\end{figure}

We present firstly the phase transition from hadronic to deconfined quark phase in  $T-\rho_B^{}$ plane
in Fig.~\ref{fig:T-RHO-with-delta-alpha=0} for symmetric matter and in Fig.~\ref{fig:T-RHO-with-delta-alpha=02}
for asymmetric matter with the global asymmetry parameter $\alpha=0.2$.

\begin{figure}[htbp]
\begin{center}
\includegraphics[scale=0.3]{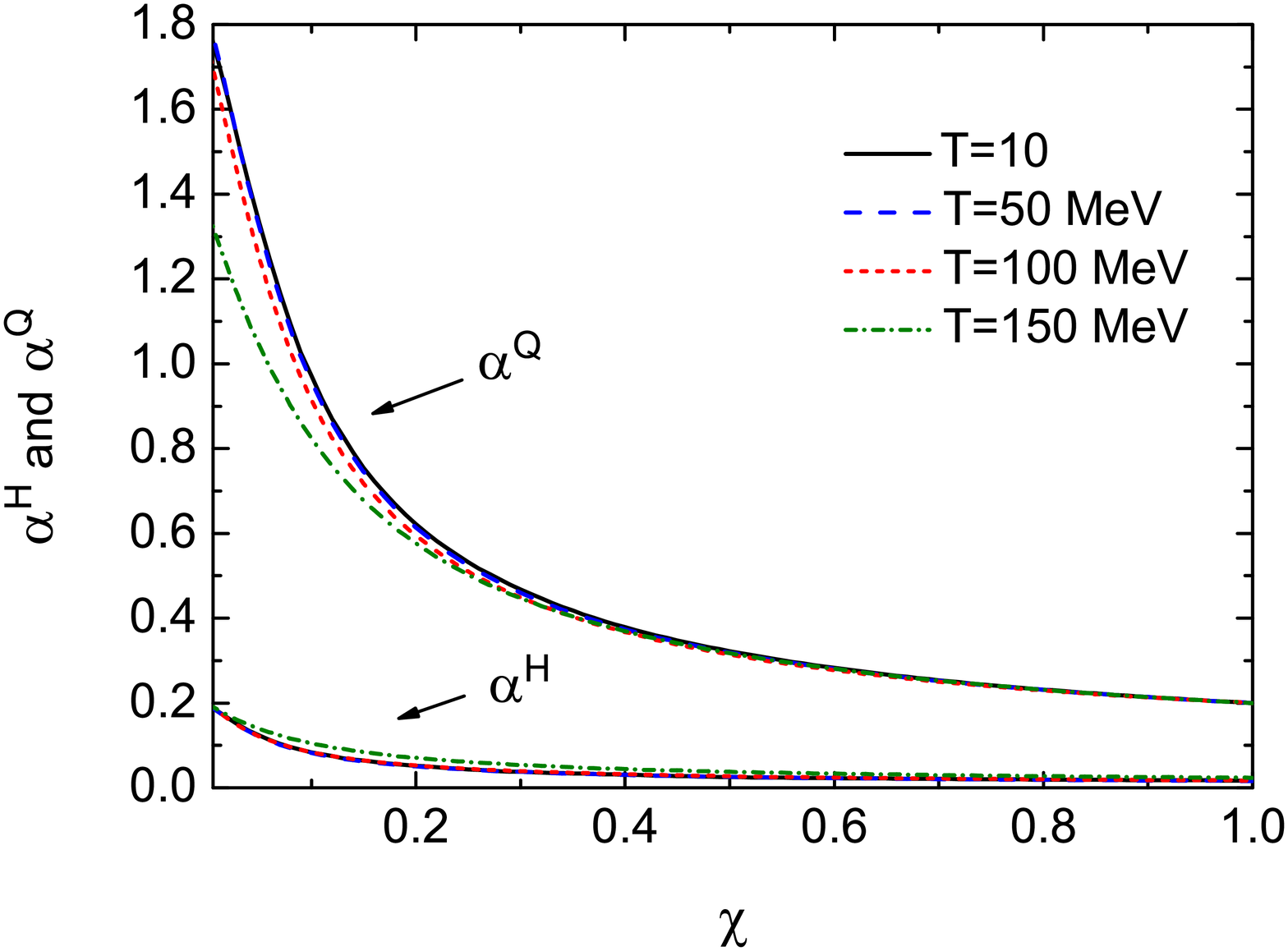}
\caption{\label{fig:kai-alphaQ-NLrho-with-delta-alpha=02} The behavior of
local asymmetric parameters $\alpha^H$
and $\alpha^Q$ in the mixed phase for several values of temperature. Parameter set
$NL\rho\delta$ is used in the calculation.}
\end{center}
\end{figure}

For symmetric matter at a fixed $T$, the first order phase transition takes place with the same pressure and $\mu_B^{}$ in both phases, but
a jump of $\rho_B^H$ to $\rho_B^Q$, just as shown in Fig.~\ref{fig:T-RHO-with-delta-alpha=0}.  In the
mixed phase, the pressures  of both phases keep unchanged and $\alpha=\alpha^H_{}=\alpha^Q_{}=0$ for any quark fraction $\chi$.
These features are quite different for the mixed phase in
isospin asymmetric matter.  As already noted in \cite{Shao11}, where the
NJL quark EoS has been used, also in the PNJL case we see a clear Isospin
Distillation effect, i.e., a strong enhancement of the isospin asymmetry in
the quark component inside the mixed phase, as reported in
Fig.~\ref{fig:kai-alphaQ-NLrho-with-delta-alpha=02}, where the asymmetry
parameter in the two components are plotted vs. the quark fraction $\chi$.
As a consequence the pressure in the mixed phase keeps rising with $\chi$,
more rapidly for quark concentrations below $50~\%$ \cite{Shao11}.

From Fig.~\ref{fig:kai-alphaQ-NLrho-with-delta-alpha=02} we remark that
this isospin enrichment of the quark phase is rather robust vs. the
increasing temperature. This is important since color pairing correlations
at low temperatures will decrease symmetry energy effects
\cite{Pagliara10}. We have to note that such large isospin distillation
effect is due to the large difference in the symmetry terms in the two phases,
mainly because all the used quark effective models do not have explicit
isovector fields in the interaction \cite{Torohq09}.

Such behavior of the local asymmetry parameters will possibly produce some
observational signals in the following hadronization during the expansion.
We can expect an inverse trend in the emission of neutron
rich clusters, as well as
 an enhancement of $\pi^-/\pi^+$, $K^0/K^+$ yield ratios from the high density
n-rich regions which undergo the transition.
Besides, an enhancement of the production of isospin-rich resonances
and subsequent decays may be found. For more details one can refer
to \cite{Torohq09, Shao11}.
Moreover, an evident feature of Fig.~\ref{fig:T-RHO-with-delta-alpha=02}
is that the onset density of hadron-quark phase
transition for asymmetric matter is much lower than
that of the symmetric one, and therefore it will be easier to  probe
in heavy-ion collision experiments.

\begin{figure}[htbp]
\begin{center}
\includegraphics[scale=0.3]{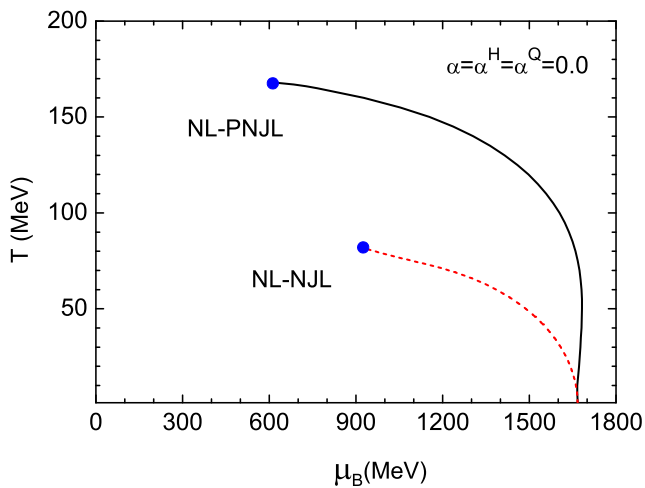}
\caption{\label{fig:T-Mu-with-delta-alpha=0}Phase diagram in $T-\mu_B^{}$ plane for symmetric matter.}
\end{center}
\end{figure}

\begin{figure}[htbp]
\begin{center}
\includegraphics[scale=0.3]{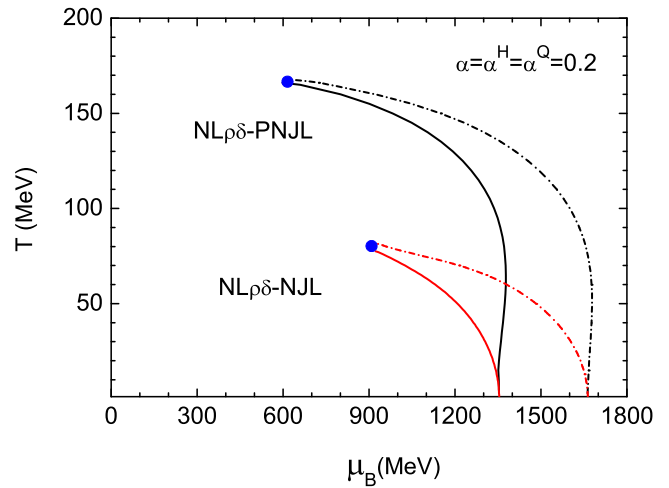}
\caption{\label{fig:T-Mu-with-delta-alpha=02}Phase diagram in  $T-\mu_B^{}$ plane for asymmetry matter with the global
asymmetry parameter $\alpha=0.2$.}
\end{center}
\end{figure}

We plot the $T-\mu_B^{}$ phase diagrams in Fig.~\ref{fig:T-Mu-with-delta-alpha=0} for  symmetric
matter and Fig.~\ref{fig:T-Mu-with-delta-alpha=02} for  asymmetric matter.
Fig.~\ref{fig:T-Mu-with-delta-alpha=0} clearly shows that there is only one phase-transition curve
in the $T-\mu_B^{}$ plane. The phase transition curve is independent of the quark fraction  $\chi$.
However, for asymmetric matter, the phase transition curve varies for different quark fraction $\chi$.
The phase transition curves  in Fig.~\ref{fig:T-Mu-with-delta-alpha=02} are obtained with
$\chi=0$ and $1$,  representing the beginning and the end of hadron-quark phase transition, respectively.

In Fig.~\ref{fig:T-Mu-with-delta-alpha=0} and Fig.~\ref{fig:T-Mu-with-delta-alpha=02}
we also plot the phase transition curves with the Hadron-NJL model. For the
NJL model with only chiral dynamics, no physical solution exists when the temperature is higher than  $\sim80$
MeV.  The corresponding temperature is enhanced to about $\sim166$ MeV with the Hadron-PNJL model, which is closer to the phase transition (crossover)
temperature given by full lattice calculation at zero or small chemical potential~\cite{Karsch01,Karsch02,Kaczmarek05}. In this sense the Hadron-PNJL model 
gives
significally different results and represents certainly an improvement
respect to the Hadron-NJL scheme of ref. \cite{Shao11}.
From Fig.~\ref{fig:T-Mu-with-delta-alpha=02} we remark that in both cases the
region around the $Critical-End-Points$ is not affected by isospin 
asymmetry contributions, which are relevant at lower temperatures and larger
chemical potentials.


From the detailed discussions of the previous two Sections now we nicely 
understand the large difference between Hadron-NJL and Hadron-PNJL phase 
transitions
and the important role of the confinement dynamics.

\begin{figure}[htbp]
\begin{center}
\includegraphics[scale=0.3]{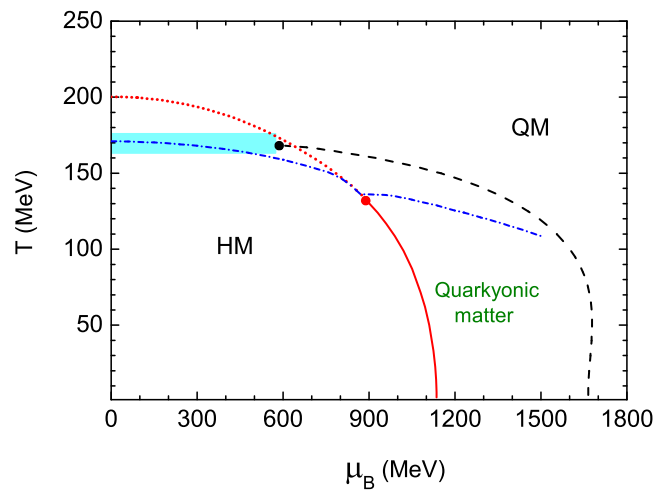}
\caption{\label{fig:T-Mu}Phase diagrams of the PNJL model and the 
Two-EoS model (dashed curve). The shaded area is just a guide for the eye.}
\end{center}
\end{figure}

Finally in Fig.~\ref{fig:T-Mu}
we present together the phase diagrams obtained by the PNJL model and the Hadron-PNJL model.
We find that the deconfined phase transition curve in the PNJL model 
is close to
that obtained in the Hadron-PNJL model at high temperature and intermediate chemical potential.
At larger chemical potential, the deconfinement phase transition curve in the PNJL model has still
a high temperature.
On the other hand from the previous Section we have seen that deconfinement 
phase transition order parameter $\Phi$
cannot describe well the phase transition at larger chemical potential and 
 lower temperatures. 
We must rely on the predictions of the Two-EoS approach, which in fact 
nicely show a good connection to the results more reliable of the PNJL 
quark model,
at high temperature and small or vanishing chemical potential.
The Two-EoS Hadron-(P)NJL model also shows
 that the phase transition at low temperature takes place at much larger 
chemical potential, consistent with the expectation of a more relevant
contribution from the hadron sector \cite{Liubo11}.

We notice that at $T=0$ there is no difference between the Hadron-NJL 
and Hadron-PNJL models. This is due to the fact that there is no dependence of 
$\Phi$ on $\mu_B$, therefore it vanishes and the PNJL reduces to the NJL.
This may casts some doubts on the reliability of the present calculations 
at $T=0$ and large $\mu_B$. However our main interest is a
region at finite T ( $T \simeq 50-100$~MeV) and $\mu_B$
( $\mu_B \simeq 1000-2000$~MeV) region that can be reached by Heavy Ion
Collisions at relativistic energies.

Moreover the results obtained by the Hadron-PNJL model at high $T$, 
small $\mu_B^{}$ and
low $T$, large $\mu_B^{}$ may be improved  with the consideration of
a stronger entanglement
between chiral condensate and Polyakov loop, and a chemical potential
dependent $T_0$  \cite{Sakai10,Sakai11}.
The relevant investigation will be performed as a further study. In any case,
 since we 
lack of reliable lattice data at large chemical potential, in general more
theoretical work is  encouraged.

\section{Summary}
In this study, the hadron-quark phase transition are investigated in 
the Two-EoS model.
The nonlinear Walecka model and the PNJL(NJL) model are used to describe
hadron matter
and quark matter, separately. We follow the  Gibbs criteria to construct the
mixed phase with baryon number and isospin conservations, likely reached
 during the hot and dense phase formed in  heavy-ion collision at
intermediate energies.
The parameters in both models are well fitted to give a good description 
of the properties of nuclear matter (even isospin asymmetric), at 
saturation as well as at higher 
baryon densities, or
lattice data at high temperature with zero/small chemical potential.

The phase diagrams for both symmetric and asymmetric matter are explored 
in both $T-\rho_B^{}$ and $T-\mu_B^{}$ planes. 
In both Hadron-(P)NJL calculations we get a first order phase transition
 with a Critical-End-Point at finite temperature and chemical potential.
In the PNJL case the $CEP$ is shifted to larger temperatures and smaller 
chemical potential, to the ($166,600$)~MeV point in the ($T, \mu_B$)
plane. This appears a nice indication of a decrease of the quark pressure
when confinement is accounted for. Such result is particularly interesting 
since the $CEP$ is now in the region of $\mu_q/T_c~\simeq~1$ (where $\mu_q$
 is the quark chemical potential) and so it could
be reached with some confidence by lattice-QCD complete calculations.

Another interesting result is that isospin effects are almost negligible
when we approach the $CEP$. At variance 
the calculation shows that the onset density
of asymmetric matter is lower than that of symmetric matter. 
Moreover in the mixed phase of asymmetric matter, the decreasing of local 
asymmetry parameter $\alpha^H$ and $\alpha^Q$
with the increasing quark fraction $\chi$ may produce some observable signals.
In particular we remark the noticeable isospin distillation mechanism 
(isospin enrichment of the quark phase) at the beginning of the mixed phase,
i.e. for low quark fractions, that should show up in the hadronization
stage during the expansion. We also see from 
 Fig.\ref{fig:kai-alphaQ-NLrho-with-delta-alpha=02} that this effect is
still there even at relatively large temperatures, certainly present in the 
high density stage of heavy ion collisions at relativistic energies 
\cite{Toro06,Toro09}. All that support the possibility of an experimental
observation 
 in the new planned facilities, for example, FAIR at GSI-Darmstadt and NICA 
at JINR-Dubna, with realistic asymmetries for stable/unstable beams. 
Some expected possible 
signals are suggested.

Because of the lack of lattice data at larger real chemical potential, 
we are left with the puzzle between chiral symmetry restoration and
deconfinement. More investigations on the chiral dynamics and (de)confinement,
  as well as their entanglement
are needed. The improvement of the understanding of quark-matter 
interaction is beneficial to get more accurate results
in the Two-EoS model.

\begin{acknowledgments}
This  project has been supported in part by the National Natural 
Science Foundation
of China under Grants Nos. 10875160, 11075037, 10935001 and the Major State
Basic Research Development Program under Contract No. G2007CB815000.
This work has been partially performed under the FIRB Research Grant
RBFR0814TT provided by the MIUR.
\end{acknowledgments}

%\end{CJK*}
\end{document}